\documentclass[prl,twocolumn,superscriptaddress,floatfix,showpacs,10pt,nofootinbib]{revtex4}%

\usepackage{latexsym}
\usepackage{amsmath,amsfonts,amsbsy,amssymb,amscd,calrsfs}
\usepackage{bbm}
\usepackage{fancybox}
\usepackage{pstricks,pst-node}
\usepackage[small,bf,hang]{caption2}
\usepackage{graphicx}
\usepackage{epsfig}
\usepackage{psfrag}
\usepackage{comment}
\usepackage[hidelinks]{hyperref}

\usepackage{float}

\psset{unit=1.3cm,linewidth=.5pt,radius=.2}  

\usepackage{multirow}                     
\usepackage{float}                          
\usepackage{lscape}                         
\usepackage{bm}

\newcommand{\bb}{\bar\beta}

\newcommand{\beq}{\begin{equation}}
\newcommand{\eeq}{\end{equation}}
\newcommand{\bi}{\begin{itemize}}
\newcommand{\ei}{\end{itemize}}
\newcommand{\bt}{\begin{tabular}}
\newcommand{\et}{\end{tabular}}
\newcommand{\bc}{\begin{center}}
\newcommand{\ec}{\end{center}}

\newcommand{\be}{\begin{equation}}
\newcommand{\ee}{\end{equation}}
\newcommand{\bea}{\begin{eqnarray}}
\newcommand{\eea}{\end{eqnarray}}
\newcommand{\ba}{\begin{array}}
\newcommand{\ea}{\end{array}}

\def\bbox{{\,\lower0.9pt\vbox{\hrule \hbox{\vrule height 0.2 cm
\hskip 0.2 cm \vrule height 0.2 cm}\hrule}\,}}
\newcommand{\dsl}{\pa \kern-0.5em /}

\font\mybb=msbm10 at 9pt
\def\bb#1{\hbox{\mybb#1}}

\def\bR {\bb{R}}
\def\bC {\bb{C}}
\def\bH {\bb{H}}
\def\bO {\bb{O}}
\def\bE {\bb{E}}

\def\bK{\bb{K}}

\def\bL {\bb{L}}
\def\bG {\bb{G}}
\def\bH {\bb{H}}

\def\bX {\bb{X}}
\def\bP {\bb{P}}

\def\bU {\bb{U}}
\def\bW {\bb{W}}
\def\bV {\bb{V}}
\def\bQ {\bb{Q}}
\def\bS {\bb{S}}

\def\tr{{\rm tr}}

\makeatletter \@addtoreset{equation}{section} \makeatother

\def\slashchar#1{\setbox0=\hbox{$#1$}           
   \dimen0=\wd0                                 
   \setbox1=\hbox{/} \dimen1=\wd1               
   \ifdim\dimen0>\dimen1                        
      \rlap{\hbox to \dimen0{\hfil/\hfil}}      
      #1                                        
   \else                                        
      \rlap{\hbox to \dimen1{\hfil$#1$\hfil}}   
      /                                         
   \fi}

\begin{document}

\title{Anti-de Sitter particles and manifest (super)isometries}

\author{Alex S. Arvanitakis}
\email{A.S.Arvanitakis@damtp.cam.ac.uk}
\affiliation{Department of Applied Mathematics and Theoretical Physics, Centre for Mathematical Sciences, University of Cambridge,
Wilberforce Road, Cambridge, CB3 0WA, U.K.}

\author{Alec E. Barns-Graham}
\email{A.E.BarnsGraham@damtp.cam.ac.uk}
\affiliation{Department of Applied Mathematics and Theoretical Physics, Centre for Mathematical Sciences, University of Cambridge,
Wilberforce Road, Cambridge, CB3 0WA, U.K.}

\author{Paul K. Townsend}
\email{P.K.Townsend@damtp.cam.ac.uk}
\affiliation{Department of Applied Mathematics and Theoretical Physics, Centre for Mathematical Sciences, University of Cambridge,
Wilberforce Road, Cambridge, CB3 0WA, U.K.}

\begin{abstract}

Starting from the classical action for a spin-zero particle in a $D$-dimensional anti-Sitter (AdS) spacetime, we recover the Breitenlohner-Freedman 
bound by quantization.  For $D=4,5,7$, and using an $Sl(2;\bK$) spinor notation for $\bK=\bR,\bC,\bH$, we find a bi-twistor form of the action
for which the AdS isometry group is linearly realised, although only for zero mass when $D=4,7$,  in agreement with previous constructions. 
For zero mass and $D=4$, the conformal isometry group is linearly  realized. We extend these results to the superparticle in the maximally supersymmetric 
``AdS$\times S$'' string/M-theory vacua, showing that  quantization yields a 128+128 component supermultiplet. We also extend them to the null string.

\end{abstract}

\maketitle

\setcounter{MaxMatrixCols}{30}

\setcounter{equation}{0}


Actions governing the dynamics of particles, strings or branes are generally invariant under the isometries, and possibly conformal isometries, 
of the background spacetime,  but these symmetries may be realized non-linearly. In some cases it is possible to make manifest the full symmetry group
 by re-expressing the action in terms of new variables that transform linearly with respect to it. 
 
A well-known example \cite{Penrose:1986ca} is the twistor formalism for massless particles in $4$-dimensional Minkowski spacetime (Mink$_4$); this  makes manifest an invariance under the $\text{Spin}(2,4)\cong SU(2,2)$  conformal isometry group of Mink$_4$ because a twistor is essentially a spinor of this group. The supertwistor \cite{Ferber:1977qx} extension of this construction to the ${\cal N}=4$ 
massless superparticle  makes  manifest the $SU(2,2|4)$ superconformal symmetry of its action \cite{Shirafuji:1983zd}, allowing a simple demonstration that its quantization yields the
${\cal N}=4$ Maxwell  supermultiplet.  Similar constructions are possible for Mink$_{3,6}$ \cite{Bengtsson:1987si}; these rely on the fact that the conformal isometry group of  
Mink$_d$ for $d=2+ {\rm dim}\, \bK$, where  $\bK=\bR,\bC,\bH$, is isomorphic to $Sp(4;\bK)$, defined as preserving a skew-$\bK$-hermitian quadratic form on $\bK^4$ \cite{Sudbery}. 

The conformal isometry group of Mink$_d$ is also the isometry group of $D$-dimensional anti-de Sitter space (AdS$_D$) for $D=d+1$. Some years ago it was 
noticed by Claus et al.  \cite{Claus:1999zh}  that the action for a particle  in AdS$_5$ could be expressed in terms of bi-twistors of Mink$_4$.  A geometric interpretation of this construction was supplied by Cederwall \cite{Cederwall:2000km}, who also showed that a similar  bi-twistor construction for AdS$_{4,7}$ could work only for zero mass.  

Here we present a simple variant of the Claus et al. construction that applies uniformly to AdS$_{4,5,7}$. Although the resulting linearly-realized $Sp(4;\bK)$ symmetry group is the AdS isometry group only for zero mass, this mismatch can be eliminated in the $\bK = \bC$ case by a redefinition of the twistor variables. We thereby recover the result of Claus et al. for  AdS$_5$,   and confirm  the conclusions of Cederwall for AdS$_{4,7}$ by algebraic means.  

Although linear realization of the  AdS$_D$ isometry group  limits our bi-twistor construction for $D=4,7$ to zero mass, a bonus for $D=4$ is that the {\it conformal} isometry group of AdS$_4$ is also linearly realized. 

Anti-de Sitter vacua arise naturally in supergravity theories. In particular the AdS$_{4,5,7}$ cases arise through the maximally supersymmetric ``AdS$\times S$'' vacua of  string/M-theory in 10/11 dimensions, 
in which context they can also be interpreted as the near-horizon geometries of, respectively,  the M2-brane, D3-brane and M5-brane \cite{Gibbons:1993sv}. The corresponding isometry supergroups
are as follows (the $O(n;\bK)$ subgroup of ${\rm OSp}(n|4;\bK)$ is defined to preserve a $\bK$-hermitian quadratic form on $\bK^n$):
\begin{equation}
\begin{array}{ccccccc} M2 &:& AdS_4 \times S^7 &:& {\rm OSp}(8|4;\mathbb{R}) \supset  {\rm Spin}(8) \times {\rm Spin}(2,3)
  \\ D3 &:& AdS_5 \times S^5 &:& {\rm OSp}(4|4;\bC) \supset {\rm U}(4) \times {\rm Spin}(2,4) 
  \\  M5 &:& AdS_7 \times S^4 &:& {\rm OSp}(2|4;\bH) \supset  {\rm USp}(4)  \times {\rm Spin}(2,6)
  \end{array} \nonumber
\end{equation}
In the D3-brane case, the AdS/CFT correspondence relates a four-dimensional $N=4$ Yang-Mills theory to IIB superstring theory in the AdS$_5 \times S^5$ backgound \cite{Maldacena:1997re}, and the superstring ground states should be described by a superparticle invariant under the ${\rm OSp}(4|4;\bC)\cong SU(2,2|4)$ isometries of this background.  

This motivates a generalization of the twistor formulation of particle dynamics in AdS to a supertwistor formulation of the superparticle. A direct construction based on AdS 
supergeometry  would involve a complicated expansion in  superspace coordinates but  a simple Mink$_d$  supersymmetrization 
suffices since the other supersymmetries  are then implied. This is reminiscent of the ``hidden'' supersymmetries of the massive superparticle \cite{Mezincescu:2014zba};  as in that case,   all supersymmetries become manifest in a supertwistor formulation, as anticipated by Cederwall \cite{Cederwall:2000km}.  
For the cases corresponding to the above table, we find that the supertwistor form of the superparticle action involves a total of $8$ fermi oscillators, so quantization will yield a supermultiplet of  $2^8=128+128$ independent states,  as expected for a maximally-supersymmetric graviton supermultiplet in the AdS$\times S$ background. 

Our constructions are  based on the fact that AdS$_D$ can be foliated by Minkowski spacetimes of dimension $d=D-1$, so it is convenient to choose coordinates
adapted to this foliation.  We will begin by showing how the  Breitenlohner-Freedman (BF) bound on the mass-squared of scalar fields in AdS \cite{Breitenlohner:1982jf}
follows from a semi-classical quantization  of the particle in such a background given that the motion on Minkowski ``slices'' is non-tachyonic.

We  start from the phase-space form of the action, invariant under reparametrizations of the particle's worldline,  which is embedded in a  $D$-dimensional 
spacetime with metric $g_{MN}$ in local coordinates $x^M$: 
\begin{equation}
S = \int \! dt \left\{ \dot x^M p_M    - \frac{1}{2} e \left(g^{MN}p_Mp_N + m^2\right)\right\}\, . 
\end{equation}
We use a ``mostly plus'' signature convention, and $e(t)$ is a Lagrange multiplier for the mass-shell constraint. 
Given an AdS$_D$ background of radius $R$, we may choose the metric to be
\begin{equation}\label{adsmet}
ds^2 = g_{MN}dx^Mdx^N = \frac{R^2}{z^2} \left(\eta_{mn}dx^mdx^n + dz^2\right)\, , 
\end{equation}
where $\{x^m; m=0,1,\dots,d-1\}$ are Minkowski coordinates for the  Mink$_d$ ``slices'', which are the
hypersurfaces of constant $z$.  AdS infinity is at $z=0$ and  there is a Killing horizon at $z=\infty$.

We can now rewrite the action as 
\begin{equation}\label{starting}
S= \int \! dt \left\{ \dot x^m p_m + \dot z p_z   -  \frac{1}{2}\tilde e \left(p^2 + \Delta^2\right)\right\}\, ,  
\end{equation}
where $R^2\tilde e = z^2 e$ and 
\begin{equation}\label{deltadef}
p^2= \eta^{mn}p_mp_n \, , \qquad \Delta^2 = p_z^2 + \left(mR/z\right)^2 \, . 
\end{equation}
Let us remark here that the physical phase space has dimension $2D-2=2d$ because the constraint also generates a gauge 
invariance, thereby lowering the dimension by $2$, and this must be the physical  phase-space dimension of any equivalent action 
in other variables.

 A feature of the action  (\ref{starting}) is that $\Delta$ is a constant of the motion. Consequently,  the motion within the $(x,p)$ subspace of phase space is that of a free particle 
of mass $\Delta$ in Mink$_d$.  The mass $m$ affects directly only the motion in the $(z,p_z)$  phase-plane.  For $m=0$ we have $\dot p_z=0$ and the motion in this phase plane
is linear.  For $m^2>0$ it is convenient to choose $\Delta>0$ and to write
\begin{equation}\label{pz0}
p_z  = \Delta\cos\varphi\, , \quad \frac{mR}{z} = \Delta \sin\varphi\, , 
\end{equation}
for angular variable $\varphi$; the motion in the $(\Delta,\varphi)$ plane is circular. 
Notice that $z=\infty$ whenever $\sin\varphi=0$, which tells us that the particle will pass through two Killing horizons
of AdS as $\varphi$ increases by $2\pi$. Because of the periodic identification of the global time coordinate of AdS and the fact that there is only one future and one past Killing horizon in one period, 
a timelike geodesic  will return to the same point in spacetime after crossing both Killing horizons. In this case we should identify $\varphi$ with $\varphi+2\pi$. 
However, a particle that crosses a Killing horizon of  the simply-connected cover of AdS will  never return to the same point in spacetime
or even the same point in space, so we should {\it not} assume that $\varphi$ is periodically identified in this case. 

We may also allow $m^2<0$ as long as $\Delta^2>0$, which implies that 
\begin{equation}
(mR)^2  > - (zp_z)^2\, . 
\end{equation}
Although $(zp_z)^2$ is non-zero on spacelike geodesics there is otherwise no classical restriction on its value, which could be zero. However, the quantum uncertainty principle  implies that its smallest value is $(\Delta z\Delta p_z)^2 = (\hbar/2)^2$. Quantum mechanics therefore implies  the inequality
\begin{equation}\label{bound}
\left(mR/\hbar\right)^2  > - \tfrac{1}{4}\, . 
\end{equation}
This is {\it not} yet a bound on the mass parameter  $M$ of the Klein-Gordon equation obeyed by the particle's wavefunction. For $m=0$ the classical action 
(\ref{starting}) is invariant under the {\it conformal} isometry group of AdS$_D$ and a quantization preserving this symmetry will yield a Klein-Gordon equation 
with mass parameter $M_c$ satisfying $(M_cR)^2=- D(D-2)/4$ \cite{Avis:1977yn}.  The Klein-Gordon  mass-parameter $M$ is therefore given by 
$M^2= M_c^2 + (m/\hbar)^2$, and the bound it satisfies is 
\begin{equation}
(MR)^2 \ge (M_cR)^2 - \tfrac{1}{4} = -d^2/4\, . 
\end{equation}
We have allowed for equality here without obvious justification;  apart from this detail, we have now recovered the BF bound for a scalar field in an AdS spacetime of arbitrary 
dimension $D=d+1$ \cite{Mezincescu:1984ev}.

This result  suggests  that we should allow all values of $m^2$ for which $\Delta^2>0$.  Of particular relevance here is the fact that in all such cases
\begin{equation}\label{keybos}
\dot z p_z = - zp_z \Delta^{-1} \dot\Delta + \tfrac{d}{dt} \left(\cdots\right)\, .
\end{equation}
Using this result, and ignoring a total derivative, we deduce that the action (\ref{starting}) is equivalent to
\begin{equation}\label{genact}
S= \int \! dt \left\{ \dot x^m p_m -  \frac{zp_z}{\Delta}\dot\Delta  -  \frac{1}{2}\tilde e \left(p^2 + \Delta^2\right)\right\}\, . 
\end{equation}
For $m=0$ we have $\Delta=p_z$. For $m^2>0$ we have $ zp_z = mR\cot\varphi$, which implies that $\varphi$ is the remaining phase space coordinate 
(and for $m=i|m|$ we have $zp_z = mR\coth\psi$ where $\Delta$ can have either sign and $\psi= -i\varphi$).

For $d=3,4,6$ we may replace the Mink$_d$ coordinates by a $2\times 2$ $\bK$-hermitian matrix $\bX$ over  $\bK = \bR,\bC,\bH$. Similarly, we may replace the 
$d$-momentum by a $2\times2$ $\bK$-hermitian matrix $\bP$ such that  $\det\bP = -p^2$ ({\it hermitian} quaternionic matrices have an intrinsically defined 
real determinant \cite{Moore,Aslaksen}). We then have
\begin{equation}
\dot x^m p_m = \tfrac{1}{4}\tr \left(\dot \bX \bP + \bP \dot\bX\right) \equiv \tfrac{1}{2}  \tr_{\mathbb{R}} (\dot\bX \bP)\, ,  
\end{equation}
where ``$\tr_{\mathbb{R}}$'' indicates the real part of the matrix trace. We now write 
\begin{equation}
\bP = \mp \bU \bU^\dagger \, , 
\end{equation}
where $\bU$ is a new $2\times 2$ matrix variable and the top/bottom sign is for positive/negative $p^0$. The mass-shell constraint is now
\begin{equation}\label{mass-shell}
\det (\bU \bU^\dagger) = \Delta^2\, . 
\end{equation}
Effectively, we have replaced the $d$-momentum by a pair of 2-component Mink$_d$ spinors, alias 2-vectors of $Sl(2;\bK)$ \cite{Kugo:1982bn}. 
This has introduced a new gauge invariance since $\bU$ is acted upon from the left by $Sl(2;\bK)$ but from the right by  \cite{Cederwall:2000km}
\begin{equation}
O(2;\bK=\bR,\bC,\bH) = O(2), \, U(2)\, , Spin(5)\, . 
\end{equation}
This ensures that $\bU$ is determined by the $d$ real variables $p_m$ up to an $O(2;\bK)$ gauge transformation. We now find that 
\begin{equation}\label{geom}
\dot x^m p_m =  \tr_\mathbb{R} \left(\dot\bU \bW^\dagger_0 \right) + \tfrac{d}{dt} \left(\cdots\right) \, , \quad  \bW_0 = \pm \bX \bU\, . 
\end{equation}
From the definition of $\bW_0$, which is also  acted upon by $Sl(2;\bK)$ from the left and by $O(2;\bK)$ from the right, it follows that
\begin{equation}\label{sshell}
\bU^\dagger \bW_0 - \bW_0^\dagger\bU \equiv 0\, . 
\end{equation}
 In the context of a particle in Mink$_{3,4,6}$ of mass $\Delta$,  we would take the Lagrangian to be $L= \tr_\mathbb{R} (\dot\bU \bW_0^\dagger)$ and impose the  identity (\ref{sshell}) as a 
 constraint with a Lagrange multiplier. The component constraints  span the Lie algebra of $O(2;\bK)$ with respect to the Poisson brackets implied by (\ref{geom}), and hence generate the required $O(2;\bK)$ gauge invariance of the action; they are the spin-shell constraints of the bi-twistor action for the massive particle in 
 Mink$_{3,4,6}$  \cite{deAzcarraga:2008ik,Routh:2015ifa,Mezincescu:2015apa} (and they also arise in other contexts, e.g.  \cite{Siegel:1994cc}). 
 Of course, in this context  we would also need to impose the new $O(2;\bK)$-invariant but $Sp(4;\bK)$-violating mass-shell constraint (\ref{mass-shell}).

 However, we are dealing with a particle in AdS$_D$ and an action (\ref{genact}) for which $\Delta$ is a phase-space  coordinate.  In this context we may interpret 
 the new mass-shell condition as providing an expression for $\Delta$ in terms of $\bU$, which is such that 
\begin{equation}
\Delta^{-1} \dot\Delta =\tr_\mathbb{R} \left(\dot\bU \bV \right)\, , \quad \bV\equiv \bU^{-1}\, . 
\end{equation}
We remark that the left and right inverses of $\bU$ are equal even for $\bK= \bH$ \cite{Zhang}. 
Taking into account (\ref{geom}), we now have 
\begin{equation}
\dot x^m p_m - \frac{zp_z}{\Delta}\dot\Delta  =  \tr_\mathbb{R} \left(\dot\bU \bW^\dagger\right) + \tfrac{d}{dt}\left(\cdots\right)\, , 
\end{equation}
where
\begin{equation}
\bW = \pm \bX \bU  -  zp_z \bV^\dagger\, .
\end{equation}
This expression for $\bW$ implies  the identity 
\begin{equation}
\bG :=  \bU^\dagger \bW - \bW^\dagger\bU \equiv 0\, , 
\end{equation}
which again becomes a constraint to be imposed by an anti-$\bK$-hermitian Lagrange multiplier $\bL$ in the action. There is no longer any mass-shell constraint, so the action is 
\begin{equation}\label{action}
S= \int \!dt\,  \tr_\mathbb{R}\left\{\dot\bU \bW^\dagger  - \bL \bG \right\} \, .
\end{equation}
There are $(3\,{\rm dim}\,\bK -2)$ first-class constraints on $8\,{\rm dim}\, \bK$ variables, yielding a physical phase space of dimension $2({\rm dim}\,\bK +2) =2d$, as required.

The $4\times2$ matrix with $\bK$-hermitian conjugate $(\bU^\dagger,\bW^\dagger)$ is  pair of Mink$_{3,4,6}$ twistors; i.e. a bi-twistor,  acted upon from the left by $Sp(4;\bK)$ and from the right by $O(2;\bK)$.  The Noether charges for the $Sp(4;\bK)$ invariance of the action (\ref{action}) are the gauge-invariant bi-twistor bilinears 
\begin{eqnarray}\label{Noether}
\mp \bU \bU^\dagger &=& \bP \, , \quad  \bU\bW^\dagger = -\bP\bX -  zp_z    \, ,\\
\pm \bW\bW^\dagger &=& -\bX\bP\bX - 2 zp_z \bX +  \left[ z^2 - (mR/\Delta)^2\right] \tilde\bP\, ,   \nonumber 
\end{eqnarray}
except that the imaginary part of $\tr(\bU\bW^\dagger)$ should be omitted for $d=4$ since this is the trace of $\bG$. The last line
uses  the mass-shell constraint (\ref{mass-shell}) and the relation
\begin{equation}
\pm \Delta^2  \bV^\dagger\bV = \tilde\bP \equiv \bP - \tr_\mathbb{R}\bP \, .  
\end{equation}
The matrix $\tilde\bP$ represents the $d$-vector $\eta^{mn}p_n$, and is such that $\det \tilde\bP =-p^2$ and
$\tr_\mathbb{R}( \bP\tilde\bP) = 2p^2$. 

For $m=0$, these Noether charges are those associated with invariance under the AdS$_D$ isometry group. In the $D=4$ case there is a larger linearly-realized symmetry 
because there is an antisymmetric second-order invariant tensor of the $SO(2)$ gauge group. Using the corresponding matrix $\bE$, and noting that  $\bU^\dagger\bW$ is
$O(2)$ invariant, we can write down an additional $4+1=5$ quadratic Noether charges: $\bU \bE \bW^\dagger$ and $\bU^\dagger\bW + \bW^\dagger \bU$. The full set of quadratic charges (omitting $\bG$ itself) spans the Lie algebra (with respect to Poisson brackets) of the AdS$_4$ conformal isometry group $SO(2,4)$. 

When $m\ne0$ the expression for $\bW\bW^\dagger$ in (\ref{Noether}) contains an additional term  that  is not  linear in momenta. This shows that the linearly realized 
$Sp(4;\bK)$ symmetry group is no longer the $Sp(4;\bK)$ isometry group (and it explains how  the action (\ref{action}) manages to be independent of the mass $m$). 
In the $\bK=\bC$ case, and $m^2>0$, this conclusion can be changed by setting
\begin{equation}\label{substitute}
\bW = \tilde\bW + i (mR) \bV^\dagger \, .
\end{equation}
Replacing $\bW\bW^\dagger$ by  $\tilde\bW\tilde\bW^\dagger$ eliminates  the unwanted $m$-dependent term in this Noether charge.  At the same time, the action in terms of  $\tilde\bW$ is unchanged from (\ref{action}) except that the $2\times 2$ anti-hermitian matrix constraint function now takes the form
\begin{equation}
\bG = \bU^\dagger \tilde \bW - \tilde\bW^\dagger\bU + 2imR\, .
\end{equation}
In other words, the $U(1)$  constraint function $\tfrac{1}{2}\tr\, \bG$ has been shifted by $2imR$, as found directly in the AdS$_5$ construction of \cite{Claus:1999zh}. 
This possibility is available only for $\bK =\bC$ because there is no imaginary unit  
for $\bK=\bR$ and a choice of one for  $\bK=\bH$ breaks the $Spin(5)$ gauge invariance. This difficulty can be circumvented by using a quartet of twistors, instead of a bi-twistor, but only at the cost of introducing second-class constraints \cite{Cederwall:2000km}.

We now return to the action (\ref{genact}) and extend its manifest Poincar\'e invariance on Mink$_d$ slices to an $N$-extended super-Poincar\'e invariance. In the $Sl(2;\bK)$ notation this is achieved by the replacement \cite{Kimura:1988ta}
\begin{equation}
\dot\bX \to \dot\bX + \sum_{i=1}^N \left(\Theta_i^\dagger \dot\Theta^i -  \dot\Theta_i^\dagger\Theta^i \right)\, , 
\end{equation}
where the $N$ {\it anticommuting} 2-component spinors $\Theta^i$ are acted upon from the left by $O(N;\bK)$ and
from the right by $Sl(2;\bK)$.  We have adopted the convention that $\bK$-conjugation (in contrast to $\bK$-hermitian conjugation) does {\it not} change the order of anticommuting
factors,  so the addition to $\dot\bX$ is hermitian. This construction ensures the existence of $N$ $Sl(2;\bK)$ spinor supercharges $\bQ^i$. 

Next, we proceed as before to the twistor form of the action, introducing the new anticommuting {\it Lorentz scalar}  variables
\begin{equation}
\Xi^i = \Theta^i \bU\, , 
\end{equation}
which are acted upon from the left by $O(N;\bK)$ and from the right by the $O(2;\bK)$ gauge group. One finds, omitting a total derivative,  that the action is 
\begin{equation}\label{saction}
S= \int \!dt\,  \tr_\mathbb{R}\left\{\dot\bU \bW^\dagger  \mp  \Xi_i^\dagger\dot\Xi^i -  \bL \bG\right\} \, ,
\end{equation}
where now
\begin{equation}
\bW = \pm \left(\bX\bU - \Theta_i^\dagger \Xi^i\right)  -  zp_z \bV^\dagger\, , 
\end{equation}
which leads to the new $O(2;\bK)$ generators
\begin{equation}
\bG= \bU^\dagger \bW - \bW^\dagger\bU  \pm  2\, \Xi_i^\dagger\Xi^i \, . 
\end{equation}

The $(4+N)\times 2$ matrix with $\bK$-hermitian conjugate $(\bU^\dagger,\bW^\dagger, \Xi_i^\dagger)$ is a  bi-supertwistor,   acted upon from the right by the $O(2;\bK)$ gauge group and 
from the left by ${\rm OSp}(N|4;\bK)$. The supersymmetry charges are  $\bQ^i =\Xi^i\bU^\dagger$  {\it and} $\bS^i= \Xi^i\bW^\dagger$, which  is double 
the number guaranteed by the construction.  In the $\bK=\bC$ case we can again allow for $m^2>0$ by making the substitution (\ref{substitute}) in the action, but now we must replace
not only the  Noether charge $\bW\bW^\dagger$ by $\tilde\bW\tilde\bW^\dagger$ but also $\bS^i$ by
\begin{equation}
\tilde\bS^i = \Xi^i\left[ \tilde \bW^\dagger - \tfrac{1}{4} \bV \, \tr\,  \bG\right] \, , 
\end{equation}
which is physically equivalent to $\Xi^i \tilde \bW^\dagger$ but the $m$-dependence of $\tilde \bW$ is cancelled by that of $\tr\, \bG$. 

Choosing $N= 8/{\rm dim}\, \bK$  we get, for $m=0$,  the invariance supergroups of the String/M-theory ``AdS$\times S$'' vacua tabulated earlier. 
In each case there are $8$ fermi oscillators so we get a supermultiplet of $2^8 =128+128$ states, which is the degeneracy of the expected graviton supermultiplet. 
In  light of the connection  between the division algebras $\bR,\bC,\bH,\bO$ and supersymmetric gauge theories in dimensions $d=3,4,6,10$ \cite{Evans:1987tm},  our results suggest that there should be some corresponding connection to the maximal  gauged supergravity theories in dimensions  $D=4,5,7$,  and perhaps $D=11$ with ``${\rm OSp}(1|4;\bO)$'' as the  AdS$_{11}$ supergroup \cite{Hasiewicz:1984ra}. 
Also, the fact that a {\it pair} of  supertwistors is needed to describe a graviton supermultiplet, whereas a single  supertwistor suffices for  a 4D Maxwell supermultiplet (to take the $\bK=\bC$ case)  could  
be viewed as support for the proposal, recently reviewed in \cite{Borsten:2015pla},  that gravity is  the  ``square'' of Yang-Mills theory.

Finally, we consider strings in AdS$_D$.  A bi-twistor action for the Nambu-Goto string  in Mink$_d$ was found in  \cite{Cederwall:1989su} but the  constraints are not all quadratic and its extension to an AdS$_D$ background is far from obvious. Here we consider the closed null string  in AdS$_{4,5,7}$.  As the twistor formulation makes manifest invariance under AdS isometries, and conformal isometries for AdS$_4$,  this may be useful for investigations  into the proposed link to higher-spin theories \cite{Sundborg:2000wp,Sezgin:2002rt,Lindstrom:2003mg}. A string-inspired  twistor model, but without spin-shell constraints,  has been used previously for this purpose \cite{Claus:1999xr}, and higher-spins emerge from 
the twistor form of the AdS (super)particle when  its spin-shell constraints are relaxed \cite{Cederwall:2000km}, but the relation of higher spin theory to the null string remains conjectural. 
  
Following the massless particle example, the standard phase-space action for the closed null string  in AdS$_D$ can be put in the form
\begin{eqnarray}
S &=& \int\! dt \oint\! d\sigma \left\{ \dot X^mP_m + \dot ZP_Z - \frac{1}{2}\tilde e \left(P^2+ P_z^2\right) \right.\nonumber \\
&&\left.  \quad - \ \ell \left(X'{}^m P_m + Z'P_Z\right) \right\} \, , 
\end{eqnarray}
where all variables are now functions of the worldsheet coordinates $(t,\sigma)$ and $\ell$ is the Lagrange multiplier for the string reparametrization constraint.
The twistor form of the action is found as before, with the result that
\begin{equation}
S = \int\! dt \oint\! d\sigma \left\{\tr_\mathbb{R}\left(\dot\bU \bW^\dagger  - \bL \bG\right) - \ell\,  \Omega\right\}\, ,
\end{equation}
where $\Omega$ is the twistor version of the string reparametrization constraint:
\begin{equation}
\Omega= \tr_\mathbb{R}  \left(\bW' \bU^\dagger - \bW^\dagger \bU' \right) \, . 
\end{equation}
This result has an obvious extension to the null $p$-brane, and supersymmetry may be incorporated  as for the particle. The zero-mode contribution
 is the bi-twistor action for the massless (super)particle.

\hfill

\noindent\textbf{Acknowledgements:}
We thank Nick Dorey for reading this manuscript, and  Martin Cederwall for many helpful discussions. A.S.A. and P.K.T. are grateful for  the hospitality and support of the Galileo Galilei Institute during the revision of this paper.  We acknowledge support from the UK Science and Technology Facilities Council (grants ST/L000385/1 and ST/M50340X/1).  A.S.A. also acknowledges support from the INFN, from Clare Hall College Cambridge, from DAMTP, from the Cambridge Philosophical society, and from the Cambridge Trust.


\providecommand{\href}[2]{#2}\begingroup\raggedright\endgroup

\end{document}